\documentclass[aps,prl,amsmath,amssymb,twocolumn,floatfix,superscriptaddress,nofootinbib]{revtex4-1}
\usepackage{graphicx}
\pdfoutput=1
\usepackage{times}
\usepackage{multirow}
\usepackage{epstopdf}
\usepackage[margin=1in]{geometry}

\usepackage{amsmath}
\usepackage{graphicx}

\begin{document}
\null\hfill\begin{tabular}[t]{l@{}}
\small{FERMILAB-PUB-19-582-ND}
\end{tabular}

\author{R.~Acciarri}
\affiliation{Fermi National Accelerator Lab, Batavia, Illinois 60510, USA}

\author{C.~Adams}
\affiliation{Argonne National Lab, Lemont, Illinois 60439, USA}

\author{J.~Asaadi}
\affiliation{University of Texas at Arlington, Arlington, Texas 76019, USA}

\author{B.~Baller}
\affiliation{Fermi National Accelerator Lab, Batavia, Illinois 60510, USA}

\author{T.~Bolton}
\affiliation{Kansas State University, Manhattan, Kansas 66506, USA}

\author{C.~Bromberg}
\affiliation{Michigan State University, East Lansing, Michigan 48824, USA}

\author{F.~Cavanna}
\affiliation{Fermi National Accelerator Lab, Batavia, Illinois 60510, USA}

\author{D.~Edmunds}
\affiliation{Michigan State University, East Lansing, Michigan 48824, USA}

\author{R.S.~Fitzpatrick}
\affiliation{University of Michigan, Ann Arbor, Michigan 48109, USA}

\author{B.~Fleming}
\affiliation{Yale University, New Haven, Connecticut 06520, USA}

\author{R.~Harnik}
\affiliation{Fermi National Accelerator Lab, Batavia, Illinois 60510, USA}

\author{C.~James}
\affiliation{Fermi National Accelerator Lab, Batavia, Illinois 60510, USA}

\author{I.~Lepetic}
\email{ilepetic@hawk.iit.edu}
\affiliation{Illinois Institute of Technology, Chicago, Illinois 60616, USA}

\author{B.R.~Littlejohn}
\affiliation{Illinois Institute of Technology, Chicago, Illinois 60616, USA}

\author{Z.~Liu}
\affiliation{Maryland Center for Fundamental Physics, Department of Physics, University of Maryland, College Park, Maryland, 20742, USA}

\author{X.~Luo}
\affiliation{University of California, Santa Barbara, CA, 93106, USA}

\author{O.~Palamara}
\email{palamara@fnal.gov}
\affiliation{Fermi National Accelerator Lab, Batavia, Illinois 60510, USA}

\author{G.~Scanavini}
\affiliation{Yale University, New Haven, Connecticut 06520, USA}

\author{M.~Soderberg}
\affiliation{Syracuse University, Syracuse, New York 13244, USA}

\author{J.~Spitz}
\affiliation{University of Michigan, Ann Arbor, Michigan 48109, USA}

\author{A.M.~Szelc}
\affiliation{University of Manchester, Manchester M13 9PL, United Kingdom}

\author{W.~Wu}
\affiliation{Fermi National Accelerator Lab, Batavia, Illinois 60510, USA}

\author{T.~Yang}
\affiliation{Fermi National Accelerator Lab, Batavia, Illinois 60510, USA}

\collaboration{The ArgoNeuT Collaboration}
\noaffiliation

\title{Improved Limits on Millicharged Particles Using the ArgoNeuT Experiment at Fermilab}

\begin{abstract}
A search for millicharged particles, a simple extension of the standard model, has been performed with the ArgoNeuT detector exposed to the Neutrinos at the Main Injector beam at Fermilab. The ArgoNeuT Liquid Argon Time Projection Chamber detector enables a search for millicharged particles through the detection of visible electron recoils. We search for an event signature with two soft hits (MeV-scale energy depositions) aligned with the upstream target. For an exposure of the detector of $1.0$ $\times$ $10^{20}$ protons on target, one candidate event has been observed, compatible with the expected background. This search is sensitive to millicharged particles with charges between $10^{-3}e$ and $10^{-1}e$ and with masses in the range from $0.1$ GeV to $3$ GeV. This measurement provides leading constraints on millicharged particles in this large unexplored parameter space region. 
\end{abstract}

\maketitle

Millicharged particles (mCPs), i.e. particles ($\chi$) with an electric charge $Q_{\chi}$=$\epsilon\cdot e$ much smaller than the  elementary  charge ($\epsilon \ll 1$), are a particularly simple, well-motivated, extension of the  standard model (SM). In their simplest form they are just new particles that violate the quantization of charge seen in the SM. They can also arise in the low-energy limit of models in which charge is quantized but there exists a kinetically mixed dark photon~\cite{Holdom:1985ag}. In addition, these particles could make up part of the dark matter in the universe~\cite{Brahm:1989jh, Boehm:2003hm, Foot:2004pa, Pospelov:2007mp, Feng:2009mn, Kaplan:2009de, Cline:2012is, Tulin:2012wi, Agrawal:2016quu}. 

Millicharged particles can be produced at any intense fixed-target-produced beam via the decays of neutral mesons or direct Drell-Yan pair production arising from proton interactions in the target
~\cite{mCPsTheory, Magill:2018tbb}~\footnote{The bremsstrahlung contribution to mCP production is not included in this study, which may further enhance the sensitivity.}. Produced mCPs are relativistic in the lab frame. For example, for a 120 GeV proton beam striking a target (as in the case of ArgoNeuT), the boost factors of the produced mCPs are in the range of 10-100. The opening angle of the mCP beam is large, of order 0.1 radians. Neutrino detectors located downstream of an intense proton beam striking a target, nominally used to produce the neutrino beam, may be exposed to a large flux of mCPs that were produced there. 
When traveling through matter, mCPs will lose energy by atomic excitation and ionization like any charged particle but with ionization and excitation rates  reduced by $\epsilon^2$. Therefore, the mCP ionization track is undetectable except when knock-on electrons energetic enough to themselves produce a visible signal are emitted. The distribution of electron recoil energies scales with the inverse squared of the electron recoil energy, 
\begin{equation}
\label{eq:firstequation}
    \frac{d\sigma}{d E_r} \simeq \frac{2\pi \alpha \epsilon^2}{m_e E_r^2}\,, 
\end{equation}
where we have taken the relativistic mCP limit.
Low-energy thresholds are therefore key to detect these ``$\delta$-rays'' produced by mCPs. 

The expected deflection of mCPs after each interaction is small.
Therefore, mCPs will travel to the detector in an approximately straight path and will point back to the target~\cite{mCPsTheory}.
Searches for mCPs have been conducted, with low-mass regions covered by low-energy experiments~\cite{Prinz:1998ua} and high-charge regions covered by collider
experiments~\cite{Vogel:2013raa,Essig:2013lka,CMS:2012xi,Jaeckel:2012yz}, 
but the mass  $m_{\chi}>$0.1~GeV and charge $Q_{\chi}<10^{-1}e$ region is unexplored.

\begin{figure}
\includegraphics[scale=0.4, trim = 1.0cm 9.0cm 1.0cm 8.5cm, clip]{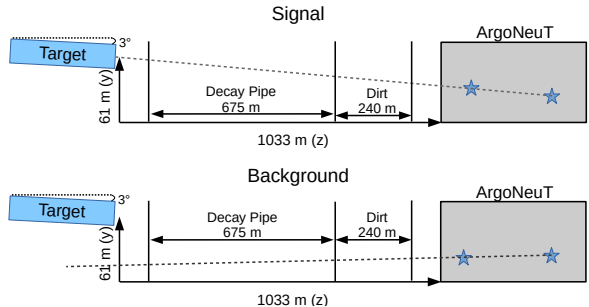}
\caption{Schematic (not to scale) of the ArgoNeuT detector location relative to the upstream NuMI target~\cite{numi}. The signal is a double-hit event with a line defined by the two hits pointing to the target (top). A background double-hit event generally will not point to the target (bottom). Figure adapted from Ref. \cite{mCPsTheory}.}
\label{fig:scheme}
\end{figure}

Liquid Argon Time Projection Chamber (LArTPC) detectors are well suited to search for these particles. As shown in~\cite{mCPsTheory}, even a short exposure of the small ArgoNeuT LArTPC detector to the Neutrinos at the Main Injector (NuMI) beam at Fermilab provides an opportunity to probe unexplored ranges of high mass ($m_\chi>0.1$~GeV) and low charge ($Q_{\chi}<10^{-1}e$). This is achieved thanks to the excellent spatial resolution and to the recently demonstrated~\cite{deExitationGamma} capability of resolving individual energy depositions down to a threshold in the sub-MeV range. These low energy depositions in LAr appear as low amplitude signals (``hits''), detected by the wire planes of the TPC. When a mCP collides with an atomic electron and the recoil electron deposits enough  energy in the LAr medium, a detectable signal (hit) is recorded by the TPC. Good background rejection is achieved by requiring two soft hits (MeV-scale energy depositions) aligned with the upstream target~\cite{mCPsTheory}, as shown schematically in Fig.~\ref{fig:scheme}. In contrast, background double-hit events will be isotropically distributed in the detector volume and will only rarely align with the target. This Letter presents the results of a search for mCPs, the first reported for a LArTPC, with the ArgoNeuT detector.

ArgoNeuT was a 0.24 ton LArTPC placed in the NuMI beam line at Fermilab for five months in 2009-2010. The TPC is $47 (w) \times 40 (h) \times 90 (l)$ cm$^3$, with two instrumented wire planes, each containing $240$ wires angled at $\pm 60$ degrees to the horizontal and spaced at $4$ mm. Signals from the wires are sampled every $198$ ns, with $2048$ samples/trigger, giving a total readout window of $405$ $\mu$s. ArgoNeuT was placed $100$ m underground in the MINOS Near Detector hall. A detailed description of the ArgoNeuT detector and its operations is given in~\cite{argoneut}. The NuMI beam~\cite{numi} is created by striking $120$ GeV protons from the Main Injector onto a graphite target.
The NuMI beam is inclined by a 3$^{\circ}$ angle with respect to ArgoNeuT. The ArgoNeuT detector was located $1033$ m downstream and $61$ m below the target (see Fig.~\ref{fig:scheme}). 

The rate of expected mCPs passing through the ArgoNeuT detector depends on the mass of the mCP. The geometrical acceptance varies between $10^{-5}$ to $10^{-7}$ for signal events~\cite{mCPsTheory}. The detection probability for double-hit signals is proportional to the fourth power of its electric charge $Q_\chi$. The detection signature of mCPs in the detector is elastic scattering with atomic electrons resulting in knock-on recoils above the detection threshold. Therefore, in order to be able to reconstruct mCPs which pass through ArgoNeuT, we search for small individual energy depositions in the detector. As recently demonstrated, in ArgoNeuT we are able to reconstruct with very good efficiency electromagnetic energy depositions as low as 300 keV~\cite{deExitationGamma}. Following the method suggested in~\cite{mCPsTheory}, to cut down on possible backgrounds in our search for mCPs we look for events with two individual soft energy depositions that are aligned with the upstream target, as shown in Fig.~\ref{fig:scheme}. 

We searched for the presence of mCPs in data from ArgoNeuT's antineutrino mode run. The trigger condition for the ArgoNeuT data acquisition was set in coincidence with the NuMI beam spill signal. A total of 4,056,940 collected triggers have been analyzed. The vast majority of NuMI beam spills delivered did not produce an observable neutrino interaction within the TPC due to the very low neutrino cross-section and the limited size of the detector, resulting in ``empty" events. In this analysis we searched for the possible presence of mCPs in these empty events. Events containing a neutrino interaction inside the LAr volume and events containing charged particles (mainly muons) produced by neutrino interactions upstream of the ArgoNeuT detector and propagating through the LArTPC volume are removed. The background for the mCP search is due to ambient gamma ray activity, beta electrons from intrinsic $^{39}$Ar activity~\cite{warp}, fluctuations of electronics noise faking signals from true energy depositions, and low-energy electrons produced by Compton scattering of photons from inelastic scattering of entering neutrons from neutrino interactions occurring upstream of the detector. To estimate the contribution due to the first three sources of background in the following we compare events acquired when the NuMI beam was operating at its typical high intensity (named ``high-beam'' in the following) to events acquired when the intensity was very low ($<1\%$ of the average intensity, named ``low-beam'' in the following). In this case the last source of background, coming from neutrino induced neutrons, is not present.

The reconstruction technique used in this analysis is described in detail in ~\cite{deExitationGamma}. It consists of a two step process, the standard LArTPC reconstruction~\cite{CC1pion} followed by a specific procedure for the identification of isolated low-energy depositions in the event. In the first stage of the analysis, hits in the recorded TPC wire signals are found, and clusters of consecutive hits are identified. Events with high-energy activity, i.e. with long tracks or showers, are removed. This leaves 3,259,427 high-beam events, corresponding to $1.0$ $\times$ $10^{20}$ protons on target (POT), and 208,730 low-beam events. 
The next step aims at efficiently identifying and reconstructing isolated low-energy activity in the selected events. Only hits localized in space within a fiducial volume region are selected, and a series of cuts is applied to possibly remove random electronics noise, as described in detail in ~\cite{deExitationGamma}. Individual signal hits whose amplitude corresponds to an energy deposition of $>300$ keV are grouped into clusters, where a cluster is defined as one or more hits on adjacent wires. For each cluster on a wire plane, we look for a corresponding cluster on the other wire plane that appears at the same time, a process called plane matching~\cite{deExitationGamma}. Plane matching keeps hits due to true energy depositions in the TPC volume and rejects hits due to electronic noise fluctuations above threshold occurring in either plane but not simultaneously in both.  Plane matching also allows for a determination of the three-dimensional (3D) position of the cluster. 

The selected clusters appear to be uniformly distributed throughout the detector volume. The average number of low-energy clusters per event is $0.15$ and $0.069$ for  high-beam events and low-beam events respectively. The vast majority of the events are empty (0 clusters) in both data sets, with a lower fraction (88\%) in the high-beam data. In the low-beam data (94\% empty events), the 1-cluster fraction ($\sim$6\%, mainly from ${}^{39}$Ar $\beta$ activity) almost exhausts the sample. The greater activity in the high-beam sample is expected to be from neutrino-produced neutrons entering the detector volume. Additional activity from low energy electrons can be anticipated to be produced by elastic interactions of mCPs generated at the neutrino beam production target.
Since our analysis method of selecting multiple soft energy depositions aligned with the upstream target is expected to be very effective in reducing the background~\cite{mCPsTheory}, we do not apply any background subtraction procedure to the data. 

\begin{figure}[t]
\includegraphics[scale=0.44, trim = 0.5cm 0.0cm 0.3cm 0.0cm, clip]{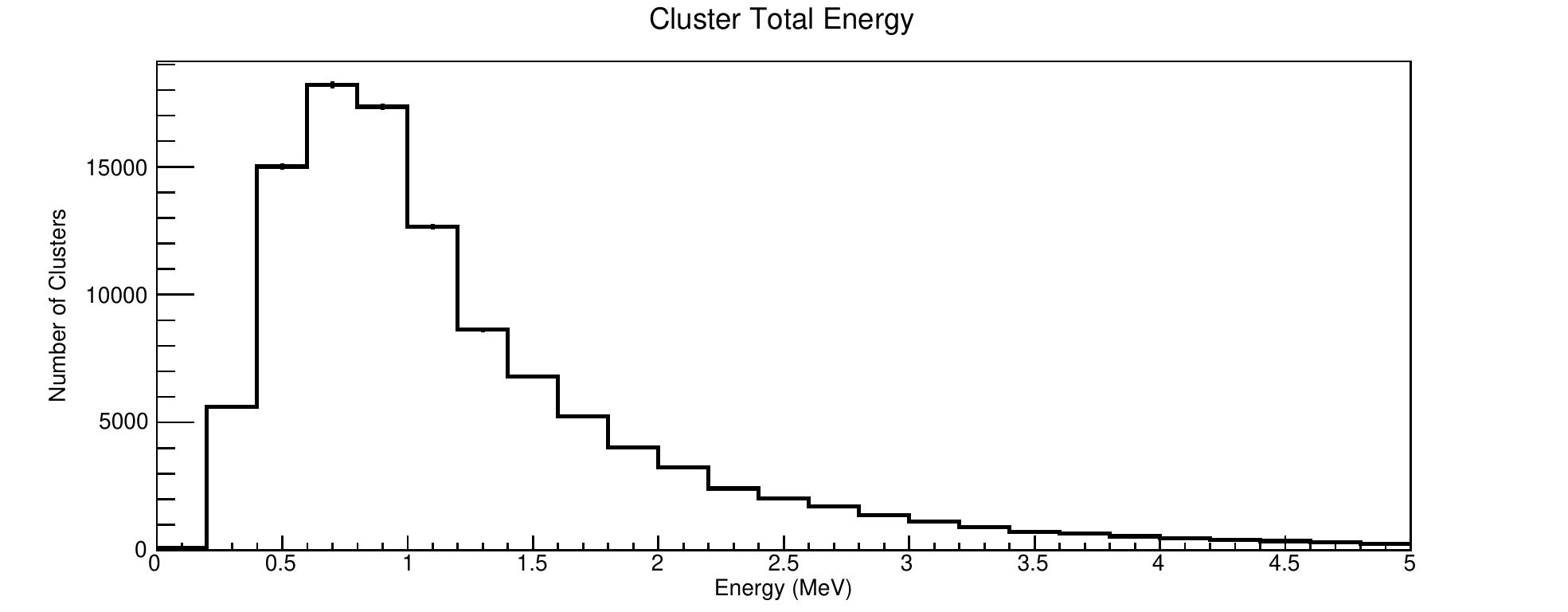}
\includegraphics[scale=0.44, trim = 0.5cm 0.0cm 0.3cm 0.0cm, clip]{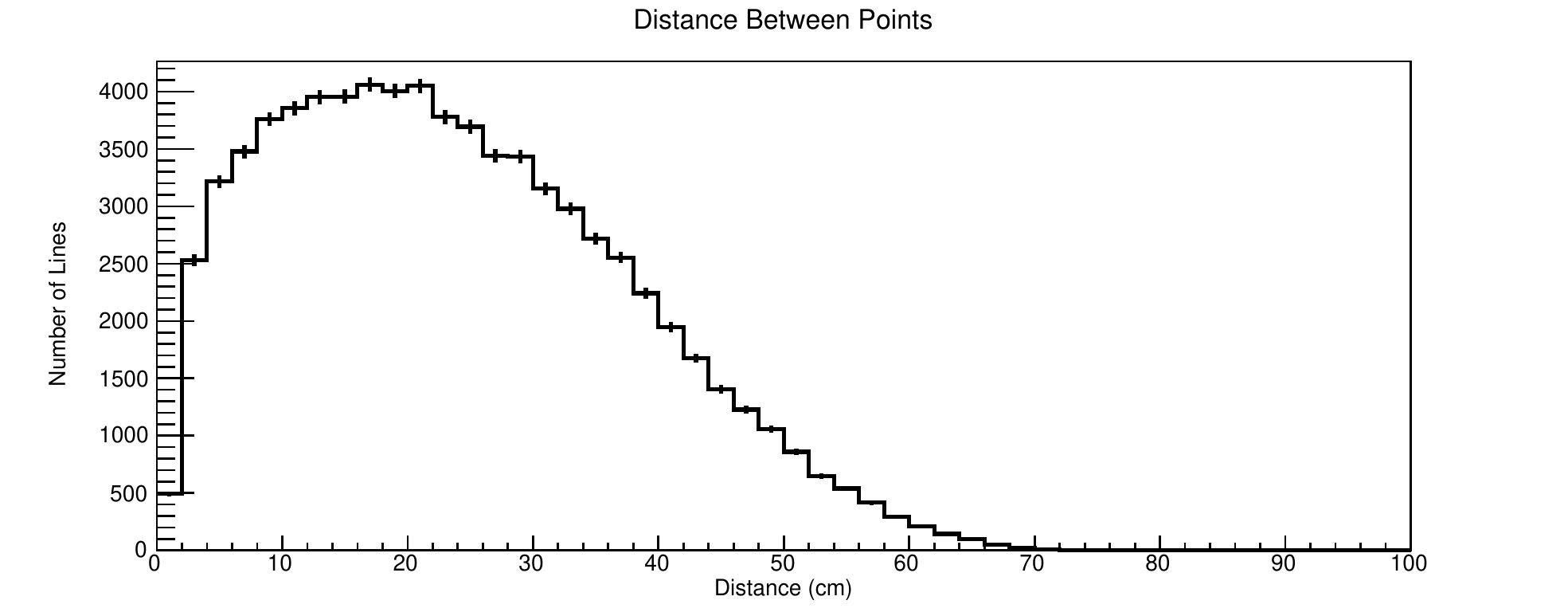}
\caption{Top: Energy deposited in each cluster in high-beam events with at least two clusters. The rising edge of the distribution is due to detector thresholding which results in a lower detection efficiency at low energies (see ~\cite{deExitationGamma}). Bottom: Distance between clusters in high-beam events with at least two clusters.}
\label{fig:TwoClusters}
\end{figure}

The final step of the analysis, the search for possible mCPs in events from the high-beam data, requires the identification of two low-energy depositions that are aligned with the upstream target (see Fig.~\ref{fig:scheme} top). The distribution of the energy deposited in each cluster and the distance between clusters for events with at least two clusters is shown in Fig.~\ref{fig:TwoClusters}. As shown in the top figure, the majority of events have energy depositions in the region around 1~MeV. For events with at least two clusters we create all possible lines that connect the two clusters. To check whether the lines point back to the target we extrapolate every line to a plane located at the downstream end of the target ($1033$ m upstream) and normal to the neutrino beam direction. The uncertainty on the location of the intersection of the line with the plane is determined by the separation of the clusters (smaller cluster spacing  corresponds to larger uncertainties) and stems from the uncertainties in the locations of the clusters inside the detector. The latter uncertainties are determined by the spatial resolution of the detector, which is $0.015$ cm in the horizontal drift direction ($x$), $0.28$ cm in the vertical direction ($y)$ and $0.16$ cm along the beam direction ($z$)~\cite{argoneut}. The smaller uncertainty in the drift direction compared to the other directions is due to the frequency of the detector readout, which samples the drift distance in $0.03$ cm samples. The uncertainties in the other two directions depend on the wire spacing and orientation of the wire planes; thus the uncertainties in the beam and vertical directions are not the same. There is also a global uncertainty of $1.52$ cm in the drift direction due to the $10 \, \mu$s beam spill window. This uncertainty in the arrival time of the beam has the same effect on both clusters in a line. While these uncertainties are small compared to the size of the detector, they can become quite large, depending on the relative location of the points, when extrapolated to the location of the target, $1033$ m upstream. Since we use the position of the intersection of the lines on the plane to identify signal events coming from the target, we want events with good directional resolution and thus place a cut of $>10$~cm on the separation between clusters. For two clusters in the center of the detector and separated by $10$ cm, the uncertainty at the target plane is about $40$ m in the vertical and $2$ m in the horizontal direction. By applying the 10~cm cut on the separation between the two clusters we are ensuring that the  uncertainties at the target plane are always smaller than these. Events where the two clusters are separated by less than 0.4 cm in the beam ($z$) direction are also ignored to remove lines with undefined slope. 

\begin{figure}
\includegraphics[scale=0.40, trim = 0.1cm 0.1cm 0.1cm 1.5cm, clip]{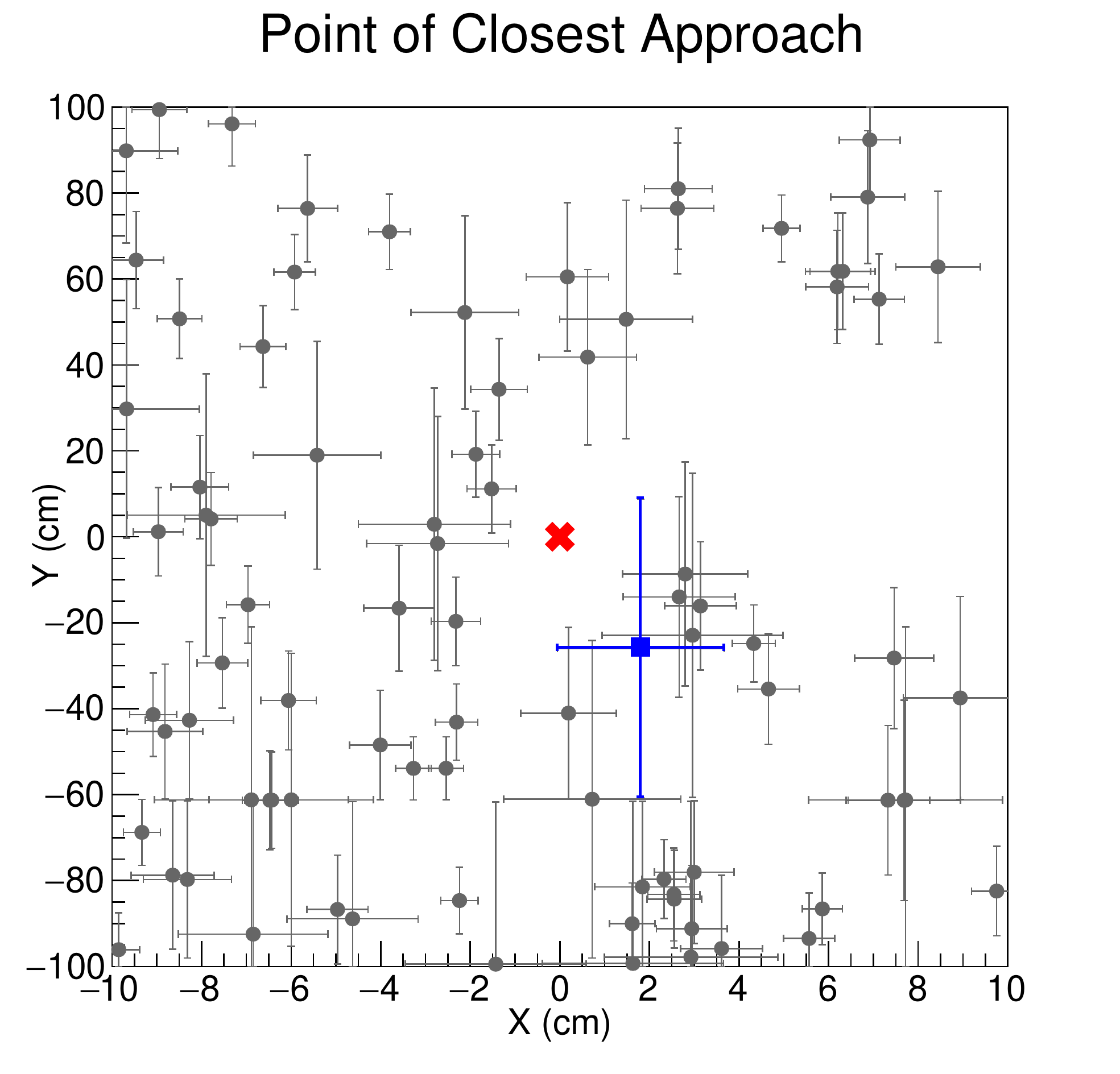}
\caption{Locations and uncertainties of the points of intersections (shown as gray circles) of lines defined by two clusters with a plane perpendicular to the beam at the downstream target's edge. The target, denoted by the red cross, is located at (0,0). The candidate signal event, denoted with a blue square, is consistent with originating from the target within its uncertainties. Only points at a distance $<$ 10 (100) m from the target in the horizontal (vertical) direction are shown.}
\label{fig:dartboard}
\end{figure}

The locations and the uncertainties of the points of intersection of the lines with the plane at the target's edge are shown in Fig.~\ref{fig:dartboard}, where the target is located at the center. 

The number of expected background events is estimated using a Monte Carlo simulation, assuming that the lines are isotropic and taking the distribution of cluster separation from data, as shown in Figure~\ref{fig:TwoClusters} (bottom). We estimate the probability that two clusters will align with the target within the uncertainties. With the detector performance parameters reported above, and taking into account the electron detection efficiency reported in Ref.~\cite{deExitationGamma}, the spatial separation of clusters and the resulting uncertainties, we expect 1.46 background events which point back to the target.  

\begin{figure}
\includegraphics[scale=0.27, trim = 0.1cm 0.01cm 0.1cm 0.5cm, clip]{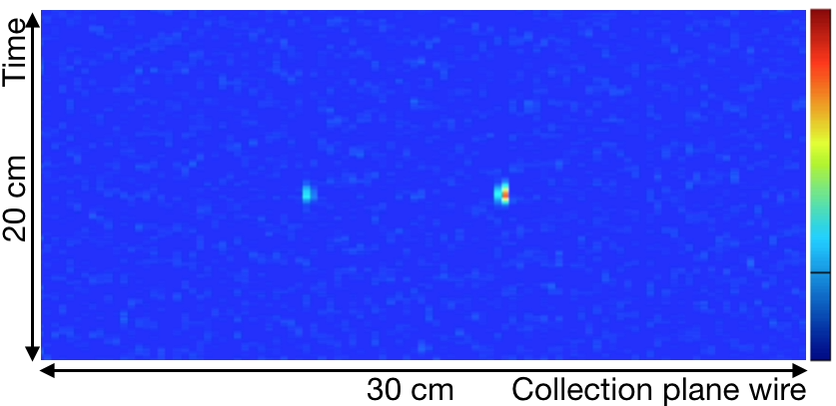}
\includegraphics[scale=0.4, trim = 0.1cm 2.0cm 0.1cm 3.0cm, clip]{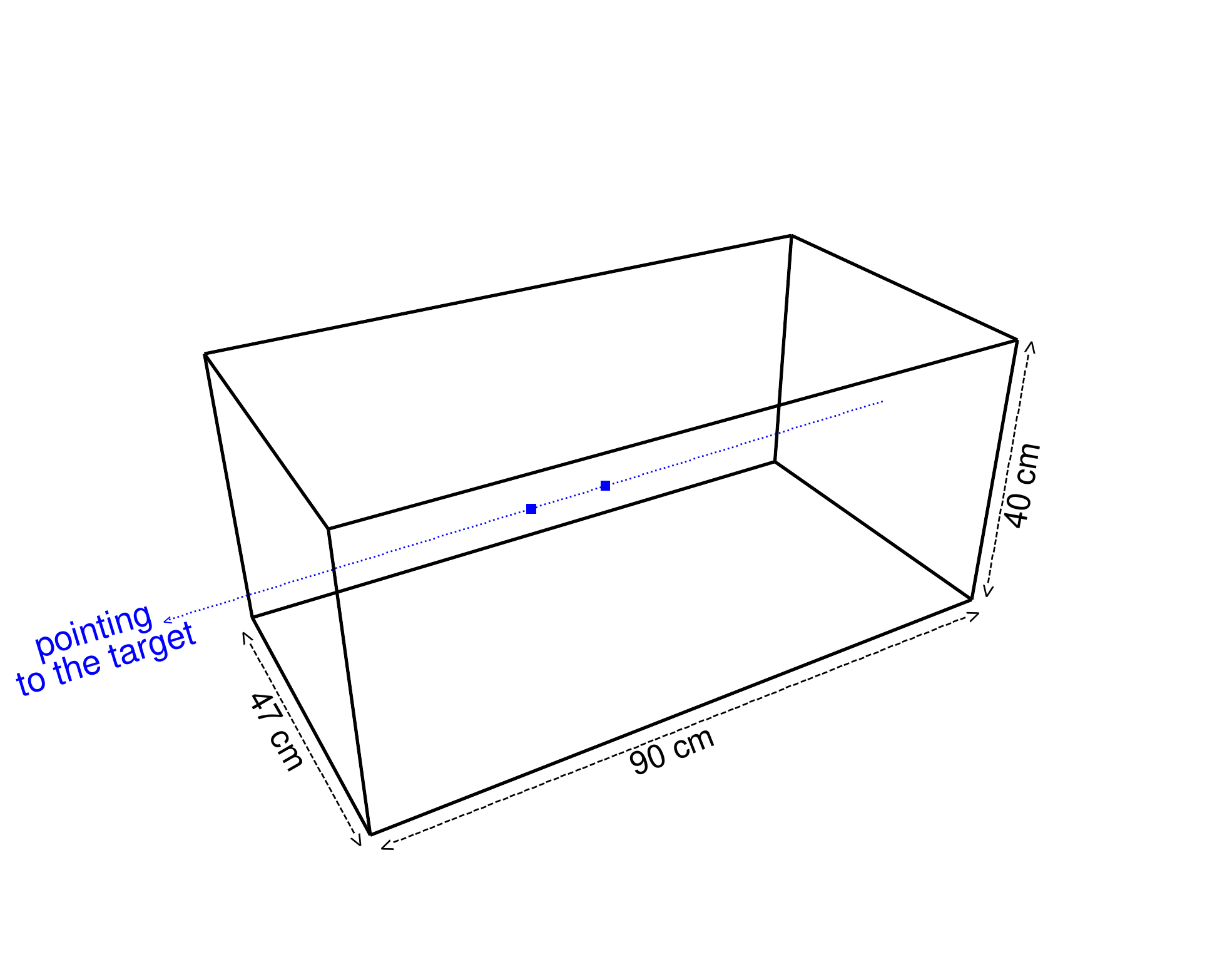}
\caption{The candidate signal event. Top: Zoomed in image from the collection wire plane. Two isolated clusters are visible in the event. Color in the image indicates the amount of charge collected. The horizontal axis is perpendicular to the collection plane wires. The vertical axis is parallel to the drift direction. Bottom: 3D reconstruction of the event with the reconstructed line superimposed.}
\label{fig:evd}
\end{figure}

We found one possible mCP signal candidate event in the ArgoNeuT data, shown as a blue square in Fig.~\ref{fig:dartboard}. The position of the line in this event overlaps with the location of the target within the horizontal and vertical uncertainties. The event, shown in Fig.~\ref{fig:evd}, has been visually scanned, and it shows no anomalies. It has two clusters spaced $11.8$ cm apart with an energy of $0.72$ ($2.82$) MeV in the more upstream (downstream) cluster. The observed candidate signal event is compatible with the expected background. 

\begin{figure}[tbp]
  \includegraphics[width=8cm]{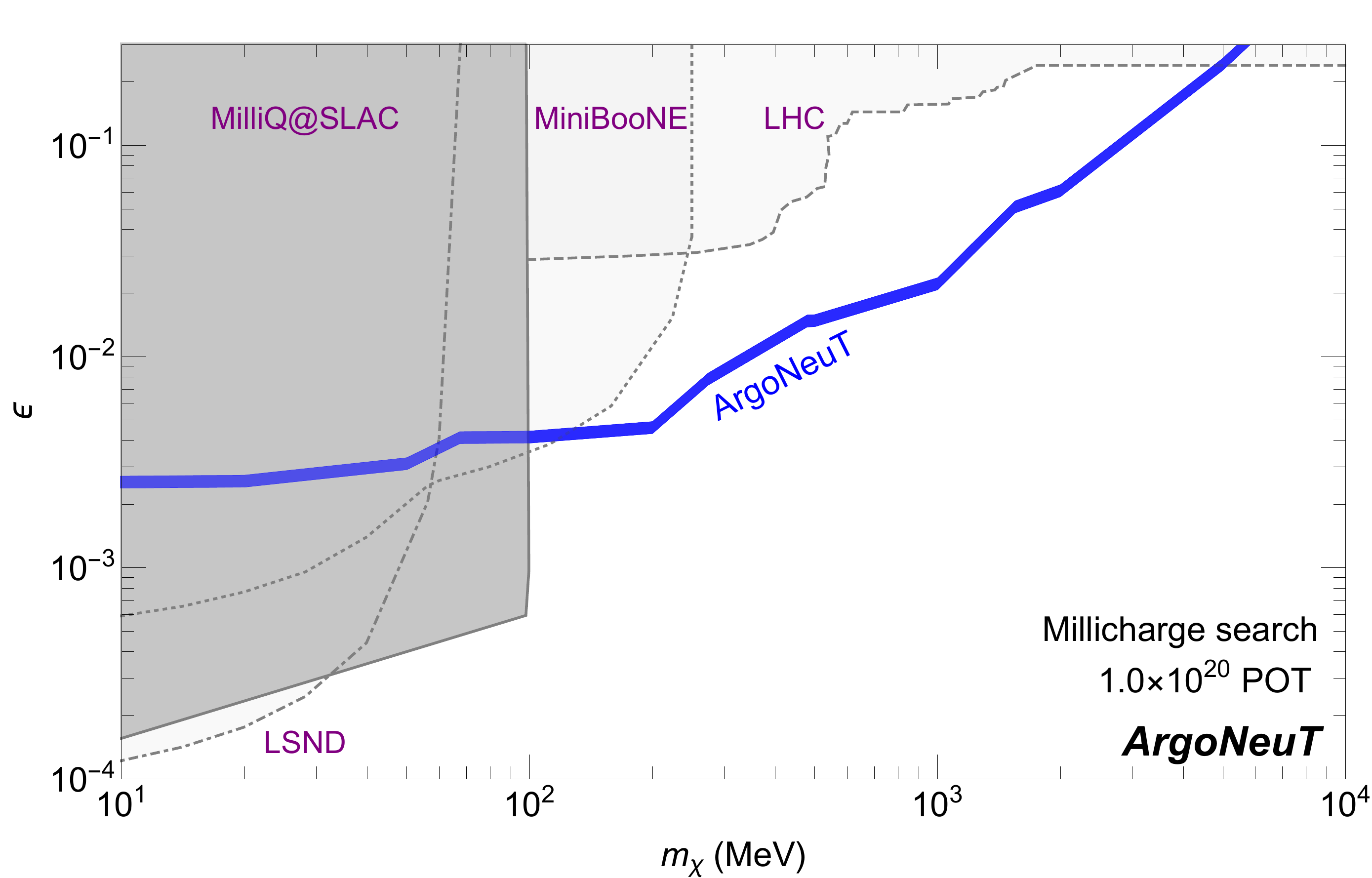}
\caption{ArgoNeuT limits (blue) in the $m_\chi-\epsilon$ plane for millicharged particles at 95\% C.L., where $\epsilon\equiv Q_\chi/e$. The limit is drawn where mCPs are unlikely to produce more than the observed number of events. The thickness of the blue band accounts for the systematic uncertainty in detector placement. 
Existing experimental limits from SLAC MilliQ~\cite{Prinz:1998ua} are shown in dark gray within a solid line. Other limits using results from the LSND and MiniBooNE neutrino experiments~\cite{Magill:2018tbb} and collider experiments~\cite{Vogel:2013raa,Essig:2013lka,CMS:2012xi,Jaeckel:2012yz} are shown in light gray within broken lines.
}
\label{fig:limit}
\end{figure}

Before using this observation to set a limit, we consider the systematic uncertainty related to ArgoNeuT's exact orientation with respect to the target. Using the spread in direction of through-going muons~\cite{muons}, we find that the direction of the target location is uncertain by $\pm 1.0^\circ$ horizontally and $\pm 0.59^\circ$ vertically. In the plane of Fig.~\ref{fig:dartboard}, this corresponds to $\pm 18$ m in $x$ and $\pm 10.6$ m in $y$. When the target location is moved within this uncertainty window,
up to five two-cluster events can be found in the ArgoNeuT high-beam data. We set limits using both one and five observed events and treat the difference as a systematic uncertainty. 
As an additional test, we have checked that the number of signal events in the plane of Fig.~\ref{fig:dartboard} is consistent with a Poisson distribution as the target location is allowed to vary across a large window (well beyond the systematic uncertainty). We have also considered the effect of the mCPs traversing the dirt en route from the target to the detector, following~\cite{mCPsTheory}. We find that the amount of energy loss is negligible in the region of interest. 
The angular deflection of an mCP from elastic scattering off of nuclei is also negligible for most of our parameter space.
The angular deflection may become of order the typical spatial resolution only for $\epsilon\sim10^{-1}$ and thus can affect the limit only for $m_\chi$ above 2 GeV. In this region, the direction from which the mCPs arrive is broadened to the order of a few meters around the target, leading to a slight increase in background. We estimate that within the systematic uncertainty, the limit is unaffected below 2 GeV and can be degraded by at most 15\% for the highest mCP masses, of order the width of the blue line in Fig.~\ref{fig:limit}.

The expected number of mCPs traversing  ArgoNeuT and their energy distribution for a given mCP mass and charge are simulated with {\tt Pythia 8}~\cite{Sjostrand:2007gs}, as detailed in Ref.~\cite{mCPsTheory}. The mean free path for every mCP is computed through equation~(\ref{eq:firstequation}) following the procedure in~\cite{mCPsTheory}, giving a probability to deposit a double hit event.
We then set limits using a CLs method~\cite{PDG} without subtracting background.
Figure~\ref{fig:limit} shows our limits on mCPs as a function of their mass and charge. We put constraints at the 95\% confidence level on mCP parameters that do not produce more than 4.7 events for one observed signal event. 
To account for the uncertainty in detector orientation discussed above, we also draw a limit on parameters that lead to more than 10.5 events, corresponding to five observed signal events, and draw a band between these two cases. We note that the limits in both these cases are very close. These upper limits on the number of expected events correspond to the conservative assumption that the background cannot be subtracted.
The results of previous experiments~\cite{Prinz:1998ua,Vogel:2013raa,Essig:2013lka,CMS:2012xi,Jaeckel:2012yz} are shown for comparison. Our result is a significant increase in the exclusion region in the range of millicharged masses $>0.1$~GeV and charge $<10^{-1}e$.

We have set new constraints from a search for millicharged particles in the ArgoNeuT LArTPC experiment at Fermilab. For a detector exposure of $1.0$ $\times$ $10^{20}$ POT, one candidate event has been observed, compatible with the expected background. ArgoNeuT has probed the region of $Q_{\chi}=10^{-1}e-10^{-3}e$ for masses in the range $m_{\chi}=0.1-3$~GeV, unexplored by previous experiments. 
This analysis represents the first search for millicharged particles in a LArTPC neutrino detector, performed with a novel search method using a cluster doublet aligned with the beam target location. The analysis techniques used in this search can be applied to future larger mass LArTPC experiments and motivate new searches.

We thank Paddy Fox for discussions.
This manuscript has been authored by Fermi Research Alliance, LLC under Contract No. DE-AC02-07CH11359 with the U.S. Department of Energy, Office of Science, Office of High Energy Physics. We wish to acknowledge the support of Fermilab, the Department of Energy, and the National Science Foundation in ArgoNeuT's construction, operation, and data analysis. We also wish to acknowledge the support of the Neutrino Physics Center (NPC) Scholar program at Fermilab, ARCS Foundation, Inc and the Science and Technology Facilities Council (STFC), part of the United Kingdom Research and Innovation Royal Society. ZL is supported in part by the NSF under Grant No. PHY1620074 and by the
Maryland Center for Fundamental Physics.

\bibliographystyle{apsrev4-1}
\bibliography{references}

\begin{thebibliography}{25}%
\makeatletter
\providecommand \@ifxundefined [1]{%
 \@ifx{#1\undefined}
}%
\providecommand \@ifnum [1]{%
 \ifnum #1\expandafter \@firstoftwo
 \else \expandafter \@secondoftwo
 \fi
}%
\providecommand \@ifx [1]{%
 \ifx #1\expandafter \@firstoftwo
 \else \expandafter \@secondoftwo
 \fi
}%
\providecommand \natexlab [1]{#1}%
\providecommand \enquote  [1]{``#1''}%
\providecommand \bibnamefont  [1]{#1}%
\providecommand \bibfnamefont [1]{#1}%
\providecommand \citenamefont [1]{#1}%
\providecommand \href@noop [0]{\@secondoftwo}%
\providecommand \href [0]{\begingroup \@sanitize@url \@href}%
\providecommand \@href[1]{\@@startlink{#1}\@@href}%
\providecommand \@@href[1]{\endgroup#1\@@endlink}%
\providecommand \@sanitize@url [0]{\catcode `\\12\catcode `\$12\catcode
  `\&12\catcode `\#12\catcode `\^12\catcode `\_12\catcode `\%12\relax}%
\providecommand \@@startlink[1]{}%
\providecommand \@@endlink[0]{}%
\providecommand \url  [0]{\begingroup\@sanitize@url \@url }%
\providecommand \@url [1]{\endgroup\@href {#1}{\urlprefix }}%
\providecommand \urlprefix  [0]{URL }%
\providecommand \Eprint [0]{\href }%
\providecommand \doibase [0]{http://dx.doi.org/}%
\providecommand \selectlanguage [0]{\@gobble}%
\providecommand \bibinfo  [0]{\@secondoftwo}%
\providecommand \bibfield  [0]{\@secondoftwo}%
\providecommand \translation [1]{[#1]}%
\providecommand \BibitemOpen [0]{}%
\providecommand \bibitemStop [0]{}%
\providecommand \bibitemNoStop [0]{.\EOS\space}%
\providecommand \EOS [0]{\spacefactor3000\relax}%
\providecommand \BibitemShut  [1]{\csname bibitem#1\endcsname}%
\let\auto@bib@innerbib\@empty
\bibitem [{\citenamefont {Holdom}(1986)}]{Holdom:1985ag}%
  \BibitemOpen
  \bibfield  {author} {\bibinfo {author} {\bibfnamefont {B.}~\bibnamefont
  {Holdom}},\ }\href {\doibase 10.1016/0370-2693(86)91377-8} {\bibfield
  {journal} {\bibinfo  {journal} {Phys. Lett.}\ }\textbf {\bibinfo {volume}
  {166B}},\ \bibinfo {pages} {196} (\bibinfo {year} {1986})}\BibitemShut
  {NoStop}%
\bibitem [{\citenamefont {Brahm}\ and\ \citenamefont
  {Hall}(1990)}]{Brahm:1989jh}%
  \BibitemOpen
  \bibfield  {author} {\bibinfo {author} {\bibfnamefont {D.~E.}\ \bibnamefont
  {Brahm}}\ and\ \bibinfo {author} {\bibfnamefont {L.~J.}\ \bibnamefont
  {Hall}},\ }\href {\doibase 10.1103/PhysRevD.41.1067} {\bibfield  {journal}
  {\bibinfo  {journal} {Phys. Rev.}\ }\textbf {\bibinfo {volume} {D41}},\
  \bibinfo {pages} {1067} (\bibinfo {year} {1990})}\BibitemShut {NoStop}%
\bibitem [{\citenamefont {Boehm}\ and\ \citenamefont
  {Fayet}(2004)}]{Boehm:2003hm}%
  \BibitemOpen
  \bibfield  {author} {\bibinfo {author} {\bibfnamefont {C.}~\bibnamefont
  {Boehm}}\ and\ \bibinfo {author} {\bibfnamefont {P.}~\bibnamefont {Fayet}},\
  }\href {\doibase 10.1016/j.nuclphysb.2004.01.015} {\bibfield  {journal}
  {\bibinfo  {journal} {Nucl. Phys.}\ }\textbf {\bibinfo {volume} {B683}},\
  \bibinfo {pages} {219} (\bibinfo {year} {2004})},\ \Eprint
  {http://arxiv.org/abs/hep-ph/0305261} {arXiv:hep-ph/0305261 [hep-ph]}
  \BibitemShut {NoStop}%
\bibitem [{\citenamefont {Foot}(2004)}]{Foot:2004pa}%
  \BibitemOpen
  \bibfield  {author} {\bibinfo {author} {\bibfnamefont {R.}~\bibnamefont
  {Foot}},\ }\href {\doibase 10.1142/S0218271804006449} {\bibfield  {journal}
  {\bibinfo  {journal} {Int. J. Mod. Phys.}\ }\textbf {\bibinfo {volume}
  {D13}},\ \bibinfo {pages} {2161} (\bibinfo {year} {2004})},\ \Eprint
  {http://arxiv.org/abs/astro-ph/0407623} {arXiv:astro-ph/0407623 [astro-ph]}
  \BibitemShut {NoStop}%
\bibitem [{\citenamefont {Pospelov}\ \emph {et~al.}(2008)\citenamefont
  {Pospelov}, \citenamefont {Ritz},\ and\ \citenamefont
  {Voloshin}}]{Pospelov:2007mp}%
  \BibitemOpen
  \bibfield  {author} {\bibinfo {author} {\bibfnamefont {M.}~\bibnamefont
  {Pospelov}}, \bibinfo {author} {\bibfnamefont {A.}~\bibnamefont {Ritz}}, \
  and\ \bibinfo {author} {\bibfnamefont {M.~B.}\ \bibnamefont {Voloshin}},\
  }\href {\doibase 10.1016/j.physletb.2008.02.052} {\bibfield  {journal}
  {\bibinfo  {journal} {Phys. Lett.}\ }\textbf {\bibinfo {volume} {B662}},\
  \bibinfo {pages} {53} (\bibinfo {year} {2008})},\ \Eprint
  {http://arxiv.org/abs/0711.4866} {arXiv:0711.4866 [hep-ph]} \BibitemShut
  {NoStop}%
\bibitem [{\citenamefont {Feng}\ \emph {et~al.}(2009)\citenamefont {Feng},
  \citenamefont {Kaplinghat}, \citenamefont {Tu},\ and\ \citenamefont
  {Yu}}]{Feng:2009mn}%
  \BibitemOpen
  \bibfield  {author} {\bibinfo {author} {\bibfnamefont {J.~L.}\ \bibnamefont
  {Feng}}, \bibinfo {author} {\bibfnamefont {M.}~\bibnamefont {Kaplinghat}},
  \bibinfo {author} {\bibfnamefont {H.}~\bibnamefont {Tu}}, \ and\ \bibinfo
  {author} {\bibfnamefont {H.-B.}\ \bibnamefont {Yu}},\ }\href {\doibase
  10.1088/1475-7516/2009/07/004} {\bibfield  {journal} {\bibinfo  {journal}
  {JCAP}\ }\textbf {\bibinfo {volume} {0907}},\ \bibinfo {pages} {004}
  (\bibinfo {year} {2009})},\ \Eprint {http://arxiv.org/abs/0905.3039}
  {arXiv:0905.3039 [hep-ph]} \BibitemShut {NoStop}%
\bibitem [{\citenamefont {Kaplan}\ \emph {et~al.}(2010)\citenamefont {Kaplan},
  \citenamefont {Krnjaic}, \citenamefont {Rehermann},\ and\ \citenamefont
  {Wells}}]{Kaplan:2009de}%
  \BibitemOpen
  \bibfield  {author} {\bibinfo {author} {\bibfnamefont {D.~E.}\ \bibnamefont
  {Kaplan}}, \bibinfo {author} {\bibfnamefont {G.~Z.}\ \bibnamefont {Krnjaic}},
  \bibinfo {author} {\bibfnamefont {K.~R.}\ \bibnamefont {Rehermann}}, \ and\
  \bibinfo {author} {\bibfnamefont {C.~M.}\ \bibnamefont {Wells}},\ }\href
  {\doibase 10.1088/1475-7516/2010/05/021} {\bibfield  {journal} {\bibinfo
  {journal} {JCAP}\ }\textbf {\bibinfo {volume} {1005}},\ \bibinfo {pages}
  {021} (\bibinfo {year} {2010})},\ \Eprint {http://arxiv.org/abs/0909.0753}
  {arXiv:0909.0753 [hep-ph]} \BibitemShut {NoStop}%
\bibitem [{\citenamefont {Cline}\ \emph {et~al.}(2012)\citenamefont {Cline},
  \citenamefont {Liu},\ and\ \citenamefont {Xue}}]{Cline:2012is}%
  \BibitemOpen
  \bibfield  {author} {\bibinfo {author} {\bibfnamefont {J.~M.}\ \bibnamefont
  {Cline}}, \bibinfo {author} {\bibfnamefont {Z.}~\bibnamefont {Liu}}, \ and\
  \bibinfo {author} {\bibfnamefont {W.}~\bibnamefont {Xue}},\ }\href {\doibase
  10.1103/PhysRevD.85.101302} {\bibfield  {journal} {\bibinfo  {journal} {Phys.
  Rev.}\ }\textbf {\bibinfo {volume} {D85}},\ \bibinfo {pages} {101302}
  (\bibinfo {year} {2012})},\ \Eprint {http://arxiv.org/abs/1201.4858}
  {arXiv:1201.4858 [hep-ph]} \BibitemShut {NoStop}%
\bibitem [{\citenamefont {Tulin}\ \emph {et~al.}(2013)\citenamefont {Tulin},
  \citenamefont {Yu},\ and\ \citenamefont {Zurek}}]{Tulin:2012wi}%
  \BibitemOpen
  \bibfield  {author} {\bibinfo {author} {\bibfnamefont {S.}~\bibnamefont
  {Tulin}}, \bibinfo {author} {\bibfnamefont {H.-B.}\ \bibnamefont {Yu}}, \
  and\ \bibinfo {author} {\bibfnamefont {K.~M.}\ \bibnamefont {Zurek}},\ }\href
  {\doibase 10.1103/PhysRevLett.110.111301} {\bibfield  {journal} {\bibinfo
  {journal} {Phys. Rev. Lett.}\ }\textbf {\bibinfo {volume} {110}},\ \bibinfo
  {pages} {111301} (\bibinfo {year} {2013})},\ \Eprint
  {http://arxiv.org/abs/1210.0900} {arXiv:1210.0900 [hep-ph]} \BibitemShut
  {NoStop}%
\bibitem [{\citenamefont {Agrawal}\ \emph {et~al.}(2017)\citenamefont
  {Agrawal}, \citenamefont {Cyr-Racine}, \citenamefont {Randall},\ and\
  \citenamefont {Scholtz}}]{Agrawal:2016quu}%
  \BibitemOpen
  \bibfield  {author} {\bibinfo {author} {\bibfnamefont {P.}~\bibnamefont
  {Agrawal}}, \bibinfo {author} {\bibfnamefont {F.-Y.}\ \bibnamefont
  {Cyr-Racine}}, \bibinfo {author} {\bibfnamefont {L.}~\bibnamefont {Randall}},
  \ and\ \bibinfo {author} {\bibfnamefont {J.}~\bibnamefont {Scholtz}},\ }\href
  {\doibase 10.1088/1475-7516/2017/05/022} {\bibfield  {journal} {\bibinfo
  {journal} {JCAP}\ }\textbf {\bibinfo {volume} {1705}},\ \bibinfo {pages}
  {022} (\bibinfo {year} {2017})},\ \Eprint {http://arxiv.org/abs/1610.04611}
  {arXiv:1610.04611 [hep-ph]} \BibitemShut {NoStop}%
\bibitem [{\citenamefont {Harnik}\ \emph {et~al.}(2019)\citenamefont {Harnik},
  \citenamefont {Liu},\ and\ \citenamefont {Palamara}}]{mCPsTheory}%
  \BibitemOpen
  \bibfield  {author} {\bibinfo {author} {\bibfnamefont {R.}~\bibnamefont
  {Harnik}}, \bibinfo {author} {\bibfnamefont {Z.}~\bibnamefont {Liu}}, \ and\
  \bibinfo {author} {\bibfnamefont {O.}~\bibnamefont {Palamara}},\ }\href
  {\doibase 10.1007/JHEP07(2019)170} {\bibfield  {journal} {\bibinfo  {journal}
  {JHEP}\ }\textbf {\bibinfo {volume} {07}},\ \bibinfo {pages} {170} (\bibinfo
  {year} {2019})},\ \Eprint {http://arxiv.org/abs/1902.03246} {arXiv:1902.03246
  [hep-ph]} \BibitemShut {NoStop}%
\bibitem [{\citenamefont {Magill}\ \emph {et~al.}(2019)\citenamefont {Magill},
  \citenamefont {Plestid}, \citenamefont {Pospelov},\ and\ \citenamefont
  {Tsai}}]{Magill:2018tbb}%
  \BibitemOpen
  \bibfield  {author} {\bibinfo {author} {\bibfnamefont {G.}~\bibnamefont
  {Magill}}, \bibinfo {author} {\bibfnamefont {R.}~\bibnamefont {Plestid}},
  \bibinfo {author} {\bibfnamefont {M.}~\bibnamefont {Pospelov}}, \ and\
  \bibinfo {author} {\bibfnamefont {Y.-D.}\ \bibnamefont {Tsai}},\ }\href
  {\doibase 10.1103/PhysRevLett.122.071801} {\bibfield  {journal} {\bibinfo
  {journal} {Phys. Rev. Lett.}\ }\textbf {\bibinfo {volume} {122}},\ \bibinfo
  {pages} {071801} (\bibinfo {year} {2019})},\ \Eprint
  {http://arxiv.org/abs/1806.03310} {arXiv:1806.03310 [hep-ph]} \BibitemShut
  {NoStop}%
\bibitem [{\citenamefont {Prinz}\ \emph {et~al.}(1998)\citenamefont {Prinz}
  \emph {et~al.}}]{Prinz:1998ua}%
  \BibitemOpen
  \bibfield  {author} {\bibinfo {author} {\bibfnamefont {A.~A.}\ \bibnamefont
  {Prinz}} \emph {et~al.},\ }\href {\doibase 10.1103/PhysRevLett.81.1175}
  {\bibfield  {journal} {\bibinfo  {journal} {Phys. Rev. Lett.}\ }\textbf
  {\bibinfo {volume} {81}},\ \bibinfo {pages} {1175} (\bibinfo {year}
  {1998})},\ \Eprint {http://arxiv.org/abs/hep-ex/9804008}
  {arXiv:hep-ex/9804008 [hep-ex]} \BibitemShut {NoStop}%
\bibitem [{\citenamefont {Vogel}\ and\ \citenamefont
  {Redondo}(2014)}]{Vogel:2013raa}%
  \BibitemOpen
  \bibfield  {author} {\bibinfo {author} {\bibfnamefont {H.}~\bibnamefont
  {Vogel}}\ and\ \bibinfo {author} {\bibfnamefont {J.}~\bibnamefont
  {Redondo}},\ }\href {\doibase 10.1088/1475-7516/2014/02/029} {\bibfield
  {journal} {\bibinfo  {journal} {JCAP}\ }\textbf {\bibinfo {volume} {1402}},\
  \bibinfo {pages} {029} (\bibinfo {year} {2014})},\ \Eprint
  {http://arxiv.org/abs/1311.2600} {arXiv:1311.2600 [hep-ph]} \BibitemShut
  {NoStop}%
\bibitem [{\citenamefont {Essig}\ \emph {et~al.}(2013)\citenamefont {Essig}
  \emph {et~al.}}]{Essig:2013lka}%
  \BibitemOpen
  \bibfield  {author} {\bibinfo {author} {\bibfnamefont {R.}~\bibnamefont
  {Essig}} \emph {et~al.},\ }in\ \href
  {http://www.slac.stanford.edu/econf/C1307292/docs/IntensityFrontier/NewLight-17.pdf}
  {\emph {\bibinfo {booktitle} {{Proceedings, 2013 Community Summer Study on
  the Future of U.S. Particle Physics: Snowmass on the Mississippi (CSS2013):
  Minneapolis, MN, USA, July 29-August 6, 2013}}}}\ (\bibinfo {year} {2013})\
  \Eprint {http://arxiv.org/abs/1311.0029} {arXiv:1311.0029 [hep-ph]}
  \BibitemShut {NoStop}%
\bibitem [{\citenamefont {Chatrchyan}\ \emph {et~al.}(2013)\citenamefont
  {Chatrchyan} \emph {et~al.}}]{CMS:2012xi}%
  \BibitemOpen
  \bibfield  {author} {\bibinfo {author} {\bibfnamefont {S.}~\bibnamefont
  {Chatrchyan}} \emph {et~al.} (\bibinfo {collaboration} {CMS Collaboration}),\
  }\href {\doibase 10.1103/PhysRevD.87.092008} {\bibfield  {journal} {\bibinfo
  {journal} {Phys. Rev.}\ }\textbf {\bibinfo {volume} {D87}},\ \bibinfo {pages}
  {092008} (\bibinfo {year} {2013})},\ \Eprint {http://arxiv.org/abs/1210.2311}
  {arXiv:1210.2311 [hep-ex]} \BibitemShut {NoStop}%
\bibitem [{\citenamefont {Jaeckel}\ \emph {et~al.}(2013)\citenamefont
  {Jaeckel}, \citenamefont {Jankowiak},\ and\ \citenamefont
  {Spannowsky}}]{Jaeckel:2012yz}%
  \BibitemOpen
  \bibfield  {author} {\bibinfo {author} {\bibfnamefont {J.}~\bibnamefont
  {Jaeckel}}, \bibinfo {author} {\bibfnamefont {M.}~\bibnamefont {Jankowiak}},
  \ and\ \bibinfo {author} {\bibfnamefont {M.}~\bibnamefont {Spannowsky}},\
  }\href {\doibase 10.1016/j.dark.2013.06.001} {\bibfield  {journal} {\bibinfo
  {journal} {Phys. Dark Univ.}\ }\textbf {\bibinfo {volume} {2}},\ \bibinfo
  {pages} {111} (\bibinfo {year} {2013})},\ \Eprint
  {http://arxiv.org/abs/1212.3620} {arXiv:1212.3620 [hep-ph]} \BibitemShut
  {NoStop}%
\bibitem [{\citenamefont {Acciarri}\ \emph {et~al.}(2019)\citenamefont
  {Acciarri} \emph {et~al.}}]{deExitationGamma}%
  \BibitemOpen
  \bibfield  {author} {\bibinfo {author} {\bibfnamefont {R.}~\bibnamefont
  {Acciarri}} \emph {et~al.} (\bibinfo {collaboration} {ArgoNeuT
  Collaboration}),\ }\href {\doibase 10.1103/PhysRevD.99.012002} {\bibfield
  {journal} {\bibinfo  {journal} {Phys. Rev.}\ }\textbf {\bibinfo {volume}
  {D99}},\ \bibinfo {pages} {012002} (\bibinfo {year} {2019})}\BibitemShut
  {NoStop}%
\bibitem [{\citenamefont {Adamson}\ \emph {et~al.}(2016)\citenamefont {Adamson}
  \emph {et~al.}}]{numi}%
  \BibitemOpen
  \bibfield  {author} {\bibinfo {author} {\bibfnamefont {P.}~\bibnamefont
  {Adamson}} \emph {et~al.},\ }\href {\doibase 10.1016/j.nima.2015.08.063}
  {\bibfield  {journal} {\bibinfo  {journal} {Nucl. Instrum. Meth.}\ }\textbf
  {\bibinfo {volume} {A806}},\ \bibinfo {pages} {279} (\bibinfo {year}
  {2016})},\ \Eprint {http://arxiv.org/abs/1507.06690} {arXiv:1507.06690
  [physics.acc-ph]} \BibitemShut {NoStop}%
\bibitem [{\citenamefont {Anderson}\ \emph
  {et~al.}(2012{\natexlab{a}})\citenamefont {Anderson} \emph
  {et~al.}}]{argoneut}%
  \BibitemOpen
  \bibfield  {author} {\bibinfo {author} {\bibfnamefont {C.}~\bibnamefont
  {Anderson}} \emph {et~al.} (\bibinfo {collaboration} {ArgoNeuT
  Collaboration}),\ }\href {\doibase 10.1088/1748-0221/7/10/P10019} {\bibfield
  {journal} {\bibinfo  {journal} {JINST}\ }\textbf {\bibinfo {volume} {7}},\
  \bibinfo {pages} {P10019} (\bibinfo {year} {2012}{\natexlab{a}})}\BibitemShut
  {NoStop}%
\bibitem [{\citenamefont {Benetti}\ \emph {et~al.}(2007)\citenamefont {Benetti}
  \emph {et~al.}}]{warp}%
  \BibitemOpen
  \bibfield  {author} {\bibinfo {author} {\bibfnamefont {P.}~\bibnamefont
  {Benetti}} \emph {et~al.} (\bibinfo {collaboration} {WARP Collaboration}),\
  }\href {\doibase 10.1016/j.nima.2007.01.106} {\bibfield  {journal} {\bibinfo
  {journal} {Nucl. Instrum. Meth.}\ }\textbf {\bibinfo {volume} {A574}},\
  \bibinfo {pages} {83} (\bibinfo {year} {2007})},\ \Eprint
  {http://arxiv.org/abs/astro-ph/0603131} {arXiv:astro-ph/0603131 [astro-ph]}
  \BibitemShut {NoStop}%
\bibitem [{\citenamefont {Acciarri}\ \emph {et~al.}(2018)\citenamefont
  {Acciarri} \emph {et~al.}}]{CC1pion}%
  \BibitemOpen
  \bibfield  {author} {\bibinfo {author} {\bibfnamefont {R.}~\bibnamefont
  {Acciarri}} \emph {et~al.} (\bibinfo {collaboration} {ArgoNeuT
  Collaboration}),\ }\href {\doibase 10.1103/PhysRevD.98.052002} {\bibfield
  {journal} {\bibinfo  {journal} {Phys. Rev.}\ }\textbf {\bibinfo {volume}
  {D98}},\ \bibinfo {pages} {052002} (\bibinfo {year} {2018})},\ \Eprint
  {http://arxiv.org/abs/1804.10294} {arXiv:1804.10294 [hep-ex]} \BibitemShut
  {NoStop}%
\bibitem [{\citenamefont {Anderson}\ \emph
  {et~al.}(2012{\natexlab{b}})\citenamefont {Anderson} \emph {et~al.}}]{muons}%
  \BibitemOpen
  \bibfield  {author} {\bibinfo {author} {\bibfnamefont {C.}~\bibnamefont
  {Anderson}} \emph {et~al.} (\bibinfo {collaboration} {ArgoNeuT
  Collaboration}),\ }\href {\doibase 10.1088/1748-0221/7/10/P10020} {\bibfield
  {journal} {\bibinfo  {journal} {JINST}\ }\textbf {\bibinfo {volume} {7}},\
  \bibinfo {pages} {P10020} (\bibinfo {year} {2012}{\natexlab{b}})},\ \Eprint
  {http://arxiv.org/abs/1205.6702} {arXiv:1205.6702 [physics.ins-det]}
  \BibitemShut {NoStop}%
\bibitem [{\citenamefont {Sjostrand}\ \emph {et~al.}(2008)\citenamefont
  {Sjostrand}, \citenamefont {Mrenna},\ and\ \citenamefont
  {Skands}}]{Sjostrand:2007gs}%
  \BibitemOpen
  \bibfield  {author} {\bibinfo {author} {\bibfnamefont {T.}~\bibnamefont
  {Sjostrand}}, \bibinfo {author} {\bibfnamefont {S.}~\bibnamefont {Mrenna}}, \
  and\ \bibinfo {author} {\bibfnamefont {P.~Z.}\ \bibnamefont {Skands}},\
  }\href {\doibase 10.1016/j.cpc.2008.01.036} {\bibfield  {journal} {\bibinfo
  {journal} {Comput. Phys. Commun.}\ }\textbf {\bibinfo {volume} {178}},\
  \bibinfo {pages} {852} (\bibinfo {year} {2008})},\ \Eprint
  {http://arxiv.org/abs/0710.3820} {arXiv:0710.3820 [hep-ph]} \BibitemShut
  {NoStop}%
\bibitem [{\citenamefont {Tanabashi~{\it et al.}}(2018)}]{PDG}%
  \BibitemOpen
  \bibfield  {author} {\bibinfo {author} {\bibfnamefont {M.}~\bibnamefont
  {Tanabashi~{\it et al.}}} (\bibinfo {collaboration} {Particle Data Group}),\
  }\href {\doibase 10.1103/PhysRevD.98.030001} {\bibfield  {journal} {\bibinfo
  {journal} {Phys. Rev. D}\ }\textbf {\bibinfo {volume} {98}},\ \bibinfo
  {pages} {030001} (\bibinfo {year} {2018})}\BibitemShut {NoStop}%
\end{thebibliography}%

\end{document}